\documentclass[pre,aps,showpacs]{revtex4}
\usepackage{graphicx}
\usepackage{amsmath}
\usepackage{amsfonts}
\usepackage{amssymb}

\begin{document}

\title{The totally asymmetric exclusion process on a ring: \\
Exact relaxation dynamics and associated model of clustering transition}

\author{J. G. Brankov $^1$, Vl. V. Papoyan $^{2,\,3}$, V. S. Poghosyan $^2$ and V. B. Priezzhev$^2$}

\affiliation{$^1$ Institute of Mechanics, Bulgarian Academy of Sciences, \\
Acad. G. Bonchev St. 4, 1113 Sofia, Bulgaria}
\email{brankov@bas.bg} \affiliation{$^2$Bogoliubov Laboratory of
Theoretical Physics, JINR, 141980 Dubna, Russia}
\affiliation{$^3$Yerevan State
University, Chair of Theoretical Physics, Alex Manoogian St. 1,
375049 Yerevan, Armenia}

\begin{abstract}
The totally asymmetric simple exclusion process in discrete time
is considered on finite rings with fixed number of particles. A
translation-invariant version of the backward-ordered sequential
update is defined for periodic boundary conditions. We prove that
the so defined update leads to a stationary state in which all
possible particle configurations have equal probabilities. Using
the exact analytical expression for the propagator, we find the
generating function for the conditional probabilities, average
velocity and diffusion constant at all stages of evolution.
An exact and explicit expression for the stationary velocity
of TASEP on rings of arbitrary size and particle filling is
derived. The evolution of small systems towards a steady state
is clearly demonstrated. Considering the generating function as
a partition function of a thermodynamic system, we study its
zeros in planes of complex fugacities. At long enough times,
the patterns of zeroes for rings with increasing size provide
evidence for a transition of the associated two-dimensional
lattice paths model into a clustered phase at low fugacities.
\end{abstract}

\maketitle


\section{Introduction}

The totally asymmetric simple exclusion process (TASEP) is,
probably, the simplest model in the kinetic theory of many
interacting particles \cite{Lig,Spohn}. Being one of the few
exactly solvable models of non-equilibrium statistical mechanics,
the TASEP describes a nontrivial evolution of the system from an
initial configuration to a steady state. While the long-time stage
of this evolution and the steady state itself have been studied in
details \cite{Dhar,Gwa,DerEvMuk,JanLeb,DerMal,DerLeb,Derrida},
much less is known about the initial short-time stage even for
relatively simple small systems.

The exact solution of the master equation obtained by Sch\"utz
\cite{Schutz} gives a complete description of the process for
arbitrary time intervals. However, the considered case of finite
number of particles on an infinite chain has a trivial steady
state in which all particles become noninteracting. By this
reason, the Sch\"{u}tz's solution was generalized for the ring
geometry \cite{Pr} where a  steady state of finite density exists.
The ring solution is valid for both continuous and discrete time
dynamics. In this paper we use the latter to demonstrate how the
system tends to its stationary state.

The discrete time TASEP, to be solvable on a ring, needs a special
translation-invariant version of the backward-ordered sequential
dynamics. The backward-ordered update is well known in both
particle-oriented and site-oriented versions \cite{Evans1,RSSS},
however, its original definitions break the invariance of the
dynamics either with respect to translations or cyclic permutations.
That is why we begin with an invariant re-definition of the
site-oriented backward-ordered sequential update for periodic boundary
conditions. Then we show that our update leads to a stationary
state where all possible particle configurations have equal
probabilities. The rest of the paper is purely computational:
using the exact formula for the conditional probabilities we find the
corresponding generating function, which is an analog of the
partition function in equilibrium statistical mechanics. From it we
obtain the evolution of the average velocity and diffusion constant. The
zeros of the generating function present a geometric portrait of the
process and give evidence for a clustering phase transition in
an associated two-dimensional lattice path problem.

\section{The model}

Since details of the particle dynamics may have a strong effect on
the time evolution of the system, we pay due attention to the
definition of an invariant backward-ordered sequential update for
the discrete-time TASEP on a ring. We consider a finite ring of
$L$ sites, each of which may be empty or occupied by a particle.
In  TASEP particles are allowed to hop only to the right: if site
$i$ is occupied and site $i+1$ is empty, the particle may hop to
site $i+1$ with probability $z$, or remain on site $i$ with
probability $1-z$. In the case of a site-oriented discrete-time
dynamics, one time-step corresponds to updating each site of the
ring in a given order. According to the standard definition of the
site-oriented backward-ordered sequential dynamics \cite{RSSS},
the labeling of the sites is fixed, $i=1,2,\ldots ,L$, ($L+i=i$),
and the local updates are applied in the order $L, L-1,\ldots ,1$.
It is readily seen that the fixed beginning of this sequence
breaks the translation invariance of the system: if site $L$ is
occupied by a particle, the possible updates depend on whether
site $1$ is empty or not. The same holds true for the
particle-oriented backward-ordered sequential dynamics
\cite{Evans1}: since particles cannot overtake each other, their
order $\mu = 1,2, \ldots, M$ is preserved in time and the local
updates always begin with particle $M$. Hence, the possible
outcomes depend on the presence of empty sites between particles
$\mu = 1$ and $\mu = M$. In the case of a factorized steady state
this feature leads to a single-particle weight for $\mu = M$
different from the (equal) weights of the other particles
\cite{Evans1}.

Our invariant, site-oriented backward-ordered sequential update is defined
as follows: at each time step one chooses an empty site, say $i$,
and begins the sequence of local updates in the order $ i, i -1,
\ldots ,i +1$. Obviously, the choice of the initial empty site
does not affect the possible outcomes of the updates in each
time step.

A distinctive property of the so defined update is the invariance
with respect to the simultaneous reversion of space and time (this
is not the case for the parallel update). To prove this, consider
a time step $t\rightarrow t+1$ from an arbitrary configuration $C$
to configuration $C^{'}$ allowed by the backward sequential
dynamics. A particle jumps from site $x$ to empty site $x+1$
with probability $z$, and stays at the same site $x$ with
probability $(1-z)$ if $x+1$ is empty, and with probability $1$
otherwise. Reversing both the time and space directions, we obtain a
step $t+1 \rightarrow t$ from $C^{'}$ to $C$. According to
the reversed dynamics, the step from $x+1$ to $x$ has probability
$z$; immobility at $x+1$ has probability $(1-z)$ if $x$ is empty
and $1$ otherwise. Noticing that the number of factors $(1-z)$ is
equal to the number of clusters of neighboring immobile particles, we see that
the total transition probabilities for the direct and
reversed dynamics coincide.

Consider now the equiprobable distribution
$P_0(C)={L\choose{P}}^{-1}$ at the initial moment of time. The
probability of a configuration $C^{'}$ at the moment $T$,
$P_T(C^{'})$ is given by
\begin{equation}
P_T(C^{'})={L\choose{P}}^{-1}\sum_{C}W_T(C\rightarrow C^{'}),
\label{forward}
\end{equation}
where $W_T(C\rightarrow C^{'})$ is the transition probability.
Continuing Eq. (\ref{forward}) and using the reversed dynamics, we
can write
\begin{equation}
P_T(C^{'})={L\choose{P}}^{-1}\sum_{C}W_T(C^{'}\rightarrow
C)={L\choose{P}}^{-1} \label{backward}
\end{equation}
due to the normalization identity
\begin{equation}
\sum_{C}W_T(C^{'}\rightarrow C)=1.
\label{normid}
\end{equation}
Therefore, if $P_0(C)$ is a stationary equiprobable distribution,
it remains equiprobable after $T$ time steps. The properties of
the transition probabilities guarantee its existence and
uniqueness. Thus, we have proved that under the above defined
backward-ordered stochastic dynamics, the system of fixed number
of particles on a finite ring, starting from arbitrary initial
configurations, evolves towards the equiprobable stationary state.

A basic object for our consideration is the conditional
probability
\begin{equation}
P_T(C|C^0)=P_T(x_1,...,x_P|x_1^0,...,x_P^0) \label{condprob}
\end{equation}
to find $P$ particles on lattice sites $0 \leq x_1 < x_2 <...<x_P
< L $ after $T$ time steps from the initial state $0 \leq x_1^0 <
x_2^0 <...<x_P^0 < L $. The Bethe ansatz decouples a complicated
many-particle dynamics into simple single-particle Bernoulli
processes characterized by the probability
\begin{equation}
\label{Bernoulli} B(N,T)= \left(
\begin{array}{c}
T\\N
\end{array}
\right) z^N(1-z)^{T-N}
\end{equation}
of $N$ advances for $T$ time steps. Here, we assume that
$B(N,T)=0$ for $N>T$. The solution obtained in \cite{Pr} reads
\begin{equation}
P_T(C|C^0)=\sum_{n_1=-\infty}^{\infty}...\sum_{n_P=-\infty}^{\infty}
(-1)^{(P-1)\sum_{i=1}^{P}n_i} \det {\bf M}, \label{solution}
\end{equation}
where the elements of the $P\times P$ matrix {\bf M} are
$$
M_{ij}=F_{s_{ij}}(x_i^0,x_j+n_jL|T)
$$
with
$$
s_{ij}=(P-1)n_j-\sum_{k\neq j}n_k+j-i,
$$
and the functions $F_m(x_i^0,x_j|T)$ are defined as
$$
F_m(x_i^0,x_j|T)=\sum_{k=0}^{\infty}\left(
\begin{array}{c}
k+m-1\\\nonumber m-1\nonumber
\end{array}
\right)B(x_i^0-k,x_j|T), \label{F0}
$$
if integer $m>0$, and
$$
F_m(x_i^0,x_j|T)=\sum_{k=0}^{-m}(-1)^k\left(
\begin{array}{c}
-m\\k
\end{array}
\right)B(x_i^0-k,x_j|T), \label{Fm}
$$
if integer $m \leq 0$. It should be noted that the infinite sums
in (\ref{solution}) are actually bounded from above and below for
finite $T$, since the arguments of non-zero Bernoulli function are
restricted by the condition $ N \leq T $.

\section{Generating function}

The probability $P_T(C|C^0)$ involves all transitions from $C^0$
to $C$ with arbitrary numbers of rotations $n_1,n_2,\ldots ,n_P$,
of the particles around the ring for time $T$. It is convenient to
select explicitly the transitions in which the total number of
steps $Y$ of all particles is fixed:
\begin{equation}
Y=L(n_1+\ldots+n_P)+(x_1+\ldots+x_P)-(x_1^0+\ldots+x_P^0).
\label{Y}
\end{equation}
The corresponding probability has the form \cite{Pr}:
\begin{equation}
\label{prob-Y}
P^Y_T(C|C^0)=\sum_{{\{n_1,\ldots,n_P\}^{'}}}(-1)^{(P-1)\sum_{i=1}^{P}n_i}\det\mathbf{M},
\end{equation}
where the summation over the set $\{n_1,\ldots,n_P\}$ is restricted by
condition (\ref{Y}).

The generating function for $P^Y_T(C|C^0)$,
\begin{equation}
\label{defgen} P^w_T(C|C^0)=\sum_{Y=0}^{\infty}w^YP^{Y}_T(C|C^0),
\end{equation}
can be written as a unrestricted weighted sum over the numbers of rotations
\begin{equation}
\label{gen} P^w_T(C|C^0)=\sum_{n_1=-\infty}^{\infty}\ldots
\sum_{n_P=-\infty}^{\infty}w^{Y}(-1)^{(P-1)\sum_{i=1}^{P}n_i}\det\mathbf{M}
\end{equation}

Summation over all possible configurations $C$ gives a generating
function $\Lambda_T(w,z)$ which can be considered as a "canonical
partition function" of the discrete process for time $T$:
\begin{equation}
\label{Gen} \Lambda_T(w,z)= \sum_{C}P^w_T(C|C^0).
\end{equation}
Actually, such a generating function has been introduced by
Derrida and Lebowitz \cite{DerLeb} in a study of the large
deviation properties of the continuous-time TASEP on a ring.

An example of $\Lambda_T(w,z)$ for the case $L=5$, $P=3$ , time
$T=3$ and initial positions $x_{1}^0=0$, $x_{2}^0=1$, $x_{3}^0=2$
is
\begin{eqnarray}
\Lambda_3(w,z)&=&1 - 3\,z + 3\,w\,z + 3\,z^2 - 12\,w\,z^2
+9\,w^2\,z^2 - z^3
+19\,w\,z^3 - 36\,w^2\,z^3 +  18\,w^3\,z^3 - 15\,w\,z^4 \nonumber\\
 &+&57\,w^2\,z^4 -63\,w^3\,z^4 + 21\,w^4\,z^4 + 6\,w\,z^5
-45\,w^2\,z^5 + 87\,w^3\,z^5 - 69\,w^4\,z^5 +21\,w^5\,z^5 -w\,z^6\nonumber\\
 &+&18\,w^2\,z^6 - 60\,w^3\,z^6 + 85\,w^4\,z^6 - 60\,w^5\,z^6
 +18\,w^6\,z^6 - 3\,w^2\,z^7  + 21\,w^3\,z^7 -48\,w^4\,z^7 \nonumber\\
 &+&60\,w^5\,z^7 - 39\,w^6\,z^7 + 9\,w^7\,z^7 - 3\,w^3\,z^8
+12\,w^4\,z^8 - 24\,w^5\,z^8 + 27\,w^6\,z^8 - 15\,w^7\,z^8 \nonumber\\
 &+&3\,w^8\,z^8 - w^4\,z^9 + 3\,w^5\,z^9 -  6\,w^6\,z^9
 +6\,w^7\,z^9 - 3\,w^8\,z^9 + w^9\,z^9.\nonumber
\end{eqnarray}
Larger polynomials can be obtained from Eqs.
(\ref{solution})-(\ref{Gen}) by using the package MATHEMATICA.
We point out that the degree of the polynomials considered here
reaches 160, and the number of digids in their integer coefficients
is up to 70.

\begin{figure}[h!]
\includegraphics[width=70mm]{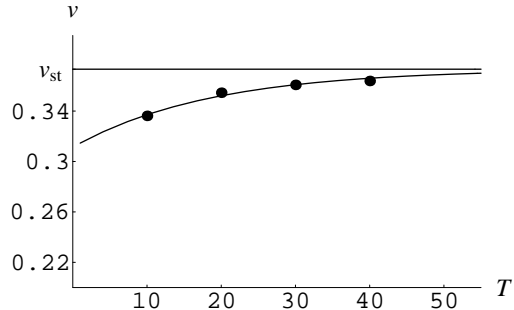}
\caption{\label{vst} Convergence of the finite-time velocity to
the stationary value for $L=8$, $P=4$.}
\end{figure}

Note that due to the probability normalization condition,
$\Lambda_T(w=1,z)\equiv 1$ for all values of the model parameters.
The negative terms in the expression for $\Lambda_T(w,z)$ appear
due to the factors $(1-z)$ attached to each cluster of immobile
particles at each time step. Using (\ref{Gen}), one easily obtains
the average distance travelled by all particles for time $T$:
\begin{equation}
\label{AverY} \langle Y\rangle=\left. \frac{\partial
{\Lambda_T(w)}}{\partial {w}} \right|_{w=1},
\end{equation}
the average velocity
\begin{eqnarray}
v=\frac{1}{P}\frac{\langle Y\rangle}{T},
\label{v}
\end{eqnarray}
and diffusion constant
\begin{eqnarray}
\Delta=\left. \frac{1}{T}\frac{\partial^2{\Lambda_T(w)}}{\partial
{w^2}}\right|_{w=1} + \frac{\langle Y\rangle-\langle
Y\rangle^{2}}{T} . \label{D}
\end{eqnarray}
The stationary velocity can be obtained from $\Lambda_T(w,z)$ for
the one-step evolution if one takes all the initial configurations with
equal weight ${L\choose{P}}^{-1}$:
\begin{equation}
v_{st}=\frac{1}{P}{L\choose{P}}^{-1}\sum_{C^0}\sum_{C} \left.
\frac{\partial}{\partial w} P_1^w(C|C^0)\right|_{w=1}. \label{vs}
\end{equation}
The above expression can be cast, see the Appendix, in the following
explicit form:
\begin{equation}
\label{vstfin}
v_{st}(L,P,z)=\frac{(P-1)!(L-P)}{(L-1)!}\sum_{i=1}^{P}\frac{(L-i-1)!}{(P-i)!}z^i=
\frac{L-P}{L-1} \;_2F_1(1,1-P,2-L;z),
\end{equation}
where $\,_2F_1$ is the hypergeometric function. Particularly, in
the thermodynamical limit (infinite chain with fixed density of
particles $\rho=P/L$) we recover the known result \cite{RSSS}
\begin{equation}
v_{st}=z \frac{1-\rho}{1- \rho z}.
\end{equation}
On the other hand, the continuous time limit yields the result
obtained by Derrida and Lebowitz \cite{DerLeb}
\begin{equation}
v_{st}=\frac{L-P}{L-1}.
\end{equation}
Several values of $v_{st}$ for small lattices $(L,P)$ at $z=0.5$ are
\begin{equation}
\label{st1} v_{st}(4,2)=\frac{5}{12} \nonumber
\end{equation}
\begin{equation}
\label{st2} v_{st}(5,2)=\frac{7}{16};\quad v_{st}(5,3)=\frac{17}{48}
\nonumber
\end{equation}
\begin{equation}
\label{st3} v_{st}(6,2)=\frac{9}{20};\quad v_{st}(6,3)=\frac{31}{80};
\quad v_{st}(6,4)=\frac{49}{160} \nonumber
\end{equation}
Convergence of velocities to their stationary values can be
obtained from (\ref{v}). Fig. \ref{vst} shows how the exactly
calculated velocities for the case $L=8$, $P=4$, $z=0.5$ and time
intervals $T=10,\, 20,\,30,\, 40$ tend to the stationary value
$v_{st}(8,4) =209/560$.

\begin{figure}[h!]
\includegraphics[width=70mm]{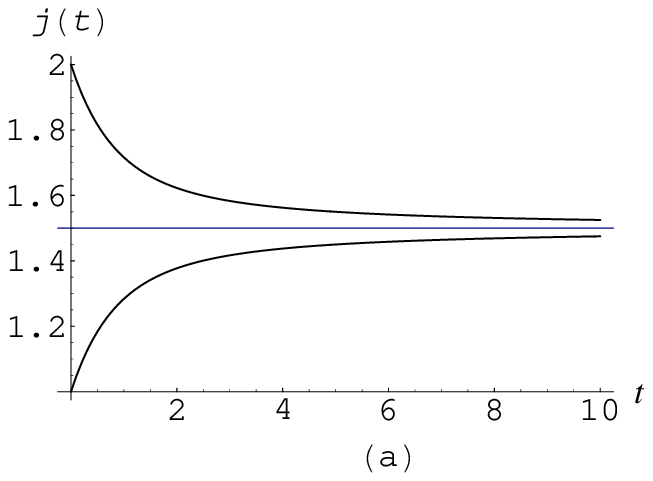}
\includegraphics[width=70mm]{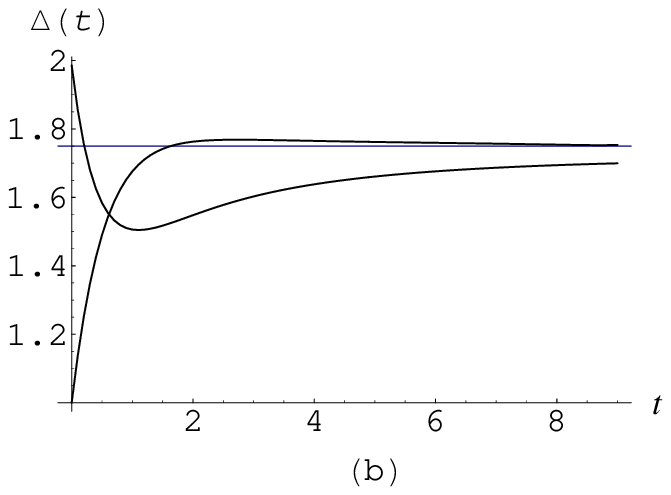}
\caption{\label{vdrlub} The tendency of current $j=Pv$ (a) and diffusion
constant (b) to stationary values (the horizontal line) for $L=5$,
$P=3$ in continuous time. The two branches correspond
to different initial conditions.}
\end{figure}

We can illustrate also the continuous-time behavior on large time scales
by rescaling the variables $z \rightarrow 0$, $T  \rightarrow \infty$,
$Tz \rightarrow t$, see Fig. \ref{vdrlub}.

As it is clearly seen, the velocity and the diffusion coefficient
relax to their known stationary values, thus giving evidence of
the correctness of the general expressions derived here. Although
the concrete way of approach to stationarity depends strongly on
the initial condition, the long-time relaxation seems to be
exponential, at least for small systems.

\section{Zeros of the partition function}

>From the equilibrium theory by Yang and Lee \cite{YL52} it is
known that the zeros of the partition function of a finite-size
system in the complex fugacity plane approach the real axis when
the system size increases. By calculating the line of zeros in the
thermodynamic limit, and their density near the real axis, one can
exactly locate the transition point and obtain the order of the
phase transition. Recently, essential progress has been made in
the extension of the Lee-Yang theory to the non-equilibrium
transitions between stationary states of the TASEP. Blythe and
Evans \cite{BE02}, \cite{BE03}, found that the distribution of the
zeros of the normalization factor for the continuous-time TASEP
with open boundaries in the plane of complex injection rate agrees
with the predictions of the Lee-Yang theory, see also \cite{BE04}.
Exact mappings of the normalization factor for the discrete-time
parallel update onto several equivalent two-dimensional lattice
path problems were obtained in \cite{BJJK}. The applicability of
the concepts of finite-size scaling and universality with respect
to the different types of updates for the open TASEP has been
established in \cite{BB}. For an overview on the recent
breakthroughs in the extension of the Lee-Yang theory to
non-equilibrium phase transitions we refer the reader to
\cite{BDL}.

\begin{figure}[h!t!]
\includegraphics[width=60mm]{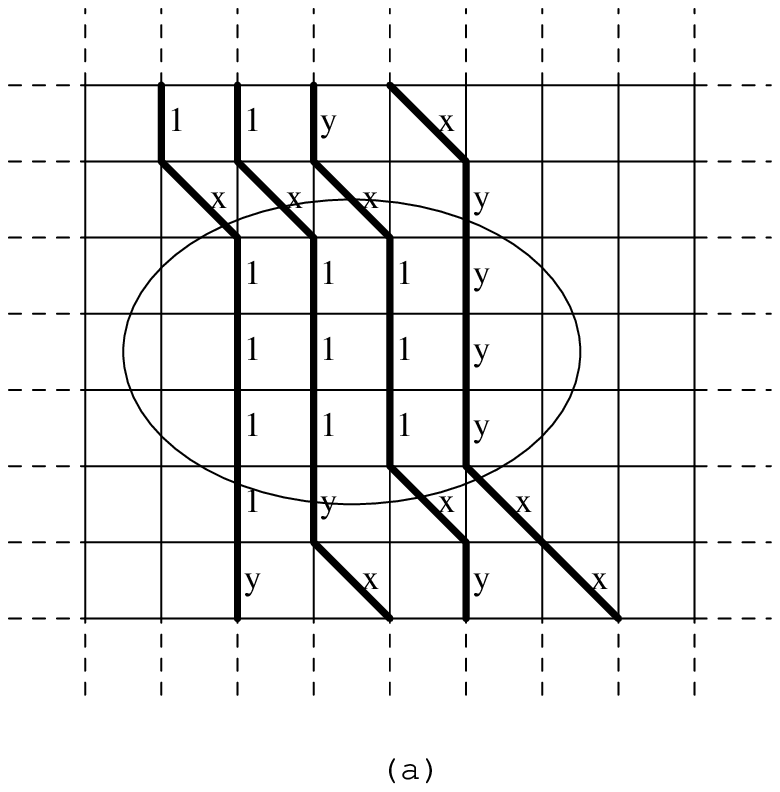}
\includegraphics[width=60mm]{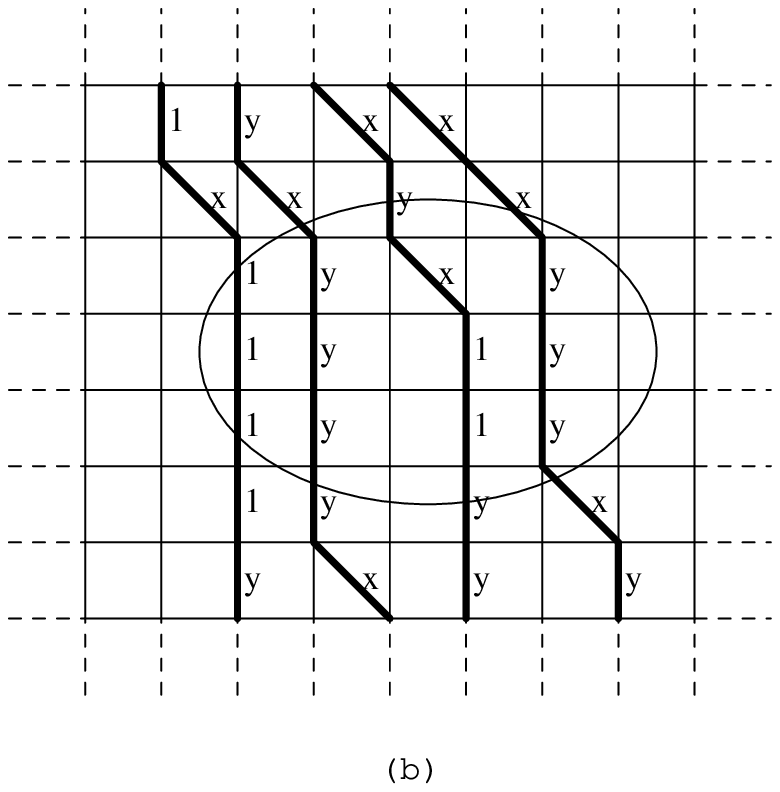}
\caption{\label{paths} Two configurations of paths are shown. The
ovals surround different clusters of particles standing during
tree time-steps: in part (a) the cluster of four standing
particles contributes a total weight of $y^3$, and in part (b) the
two clusters, of two standing particles each, contribute a total
weight of $xy^6$.}
\end{figure}

\begin{figure}[h]
\includegraphics[width=80mm]{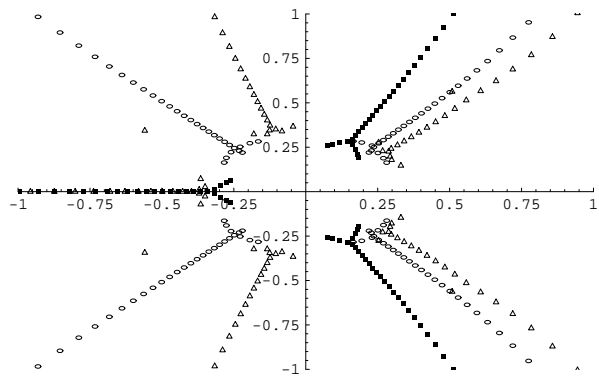}
\caption{\label{zerosx} The pattern of zeros of the partition
function $Z_{T}(x,0.5)$ in the complex-$x$ plane for $L=6$, $P=3$,
$T=40$ (solid squares), $L=8$, $P=4$, $T=40$ (open circles),
and $L=10$, $P=5$, $T=30$ (open triangles). Note that the number
of branches equals the number of particles $P$.}
\end{figure}

\begin{figure}[t]
\includegraphics[width=70mm]{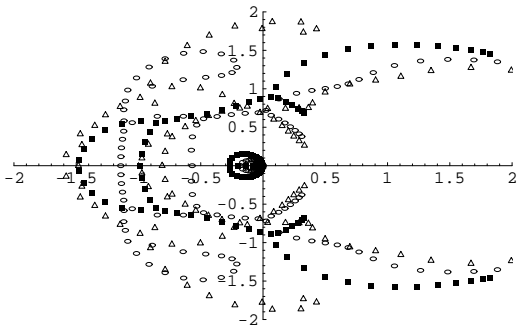}
\caption{\label{zerosy} The pattern of zeros of the partition
function $Z_{T}(0.5,y)$ in the complex-$y$ plane for $L=6$, $P=3$,
$T=40$ (solid squares), $L=8$, $P=4$, $T=40$ (open circles),
and $L=10$, $P=5$, $T=30$ (open triangles).}
\end{figure}

Let us now re-interpret the generating function (\ref{Gen}) as a
partition function of a two-dimensional lattice path problem, in
which the horizontal coordinates correspond to the original
lattice sites, and the vertical direction is the discrete time.
Each step to the right has a statistical weight $x=wz$, and each
vertical step has a weight $y=(1-z)$ if the target site for the
next step is empty and the weight 1 if it is occupied. Fig.
\ref{paths} illustrates the equivalent two-dimensional path
problem. The statistical weights $x,y,1$ are ascribed to all steps
of the paths in accordance with the TASEP dynamics. In terms of
these new variables, (\ref{Gen}) takes the form of a polynomial
$Z_{T}(x,y)$ with positive coefficients. For instance, the
polynomial corresponding to $\Lambda_3(w,z)$ is
\begin{eqnarray}
\label{NewGen} Z_3(x,y)&=&x^9 + 3x^8 y + 3x^7 y(1+2y) + 3x^6
y(1+3y+2y^2)+ 3x^5 y^2(2+4y+y^2)+ x^4 y^2(3+10y+7y^2+y^3)\nonumber
\\&+& 3x^3 y^2(1+2y+2y^2+y^3) + 3x^2
y^3(1+y+y^2)+ x y^3(1+y+y^2)+y^3.
\end{eqnarray}
Obviously, the probability normalization condition now implies
$Z_T(x,1-x) \equiv 1$. Inspired by the success of the Lee-Yang
theory in explaining how singularities may build up in the
thermodynamic limit, we attempt to clarify the analytical
structure of $Z_{T}(x,y)$ by studying the location of its zeros in
the planes of complex fugacities $x$ and $y$, for sufficiently
long times $T$, different ring sizes $L$ at fixed particle density
$P/L$ and given initial conditions. The patterns of zeros obtained
in the complex-$x$ and complex-$y$ planes are illustrated in Figs.
\ref{zerosx} and \ref{zerosy}, respectively, for three cases: $L=6$,
$P=3$, $T=40$; $L=8$, $P=4$, $T=40$; and $L=10$, $P=5$, $T=30$.

By inspecting these figures, one may speculate on the tendency of
the zeros to approach the positive fugacity axis with increasing
$L$ at fixed $P/L$. That may signal the appearance of a kind of
condensation phase transition in the related thermodynamic model
with fugacities $x$ and $y$. Arguments in favor of such a
conjecture are given in the next section.

\section{Condensation}

To reveal the nature of the possible phase transition, it is
convenient to use the mapping of the TASEP onto the zero-range
process (ZRP), see \cite{ZRP} for a recent review. Consider a ZRP
each site of which is identified, in the same sequential order,
with one of the $L-P$ empty sites of the TASEP. Let the number of
particles at each site of the ZRP be equal to the number of
neighboring occupied sites of the TASEP just on the left of the
corresponding empty site.

The interpretation of the partition function $Z_T(x,y)$ in terms
of the ZRP is the following. Each jump of a particle from site $i$
to site $i+1$ $(i=1,2,\dots, L-P; i+L-P=i)$ for the unit time step
from $t$ to $t+1$ carries a weight $x$, as in the TASEP. Each site
occupied by least one particle which does not move during that
time step has the weight $y$. The total contribution to $Z_T(x,y)$
from a particular realization of the process is the product of
weights over all time steps from $0$ to $T$. Comparing the
contributions with different factors $y^n$ and $y^m$, $n \leq m$,
and equal number of spatial steps (i.e. factors $x^k$), we see
that the decrease in $y$ leads to reduction in the average number
of sites occupied by immobile particles and, therefore, to an
effective clustering attraction between them. For sufficiently
small $y$, i.e. sufficiently strong attraction, a condensation of
particles may take place at a site (or at a number of sites).

Real-space condensation phenomena are known to take place in the
stationary states of both the heterogeneous ZRP, with
site-dependent hopping probabilities, and the homogeneous ZRP,
where the hopping rates depend only on the occupation number of
the departure site \cite{ZRP}. In the former case the $P$
particles of the ZRP are viewed as bosons which may have $N=L-P$
energies determined by the different single-site hopping rates;
the energy levels are defined so that the ground state corresponds
to the lowest hopping rate. In the discrete-time TASEP with
quenched random hopping probabilities of the particles
\cite{Evans1}, the phase transition takes place between a
low-density inhomogeneous phase, with a traffic jam behind the
slowest particle (and empty space in front of it), and a
high-density congested phase.

In our case the condensation is due to a specific interplay
between the immobility probability $y$ and the fugacity $x$
associated with the hopping probability. Turning back to the
effective two-dimensional path problem, we see that vertical
bonds, corresponding to immobile neighboring particles, tend to
form compact clusters, or jams in the TASEP traffic. This
situation is shown in the ovals in Fig. \ref{paths}. The cluster
of four standing particles in Fig. \ref{paths} (a) lives three
time steps and brings the total weight $y^3$. The configuration of
paths in Fig. \ref{paths} (b) corresponds to two clusters
containing two standing particles each. The weight of the
configuration in the second oval is $xy^6$ that is much less than
$y^3$ for small $y$. Varying parameters $x,y$, one may attain an
arbitrary strong effective attraction between clusters.

\section{Discussion}

Here we considered the exact evolution of the TASEP on small rings
during finite intervals of discrete time. The results for the
velocity and diffusion constant are not surprising. As expected,
both of them tend to their stationary values exponentially,
although the particular features of the convergence depend on the
initial conditions.

A more interesting object is the generating function of the
process. Considering it as a function of two independent
variables, fugacities of immobile particles and particle jumps, we
studied the loci of its zeros in the planes of complex fugacities.
When the location of the zeros in the right-hand half-plane is
well behaved, a finite-size scaling analysis can provide important
results concerning the very existence, the location, and the
characteristics of the conjectured phase transition. However, the
study of larger lattices and longer evolution times poses serious
computational problems and will be presented in a forthcoming
work. Although the tractable system sizes are too small, the
obtained patterns suggest that with increasing the lattice size at
a fixed particle density, the zeros may concentrate at a complex
phase boundary which cuts the positive real axis at the transition
point. When the corresponding fugacity crosses that point, a phase
transition in the related two-dimensional equilibrium lattice path
model should occur. We interpret  this transition as a
condensation of immobile particles. However, several problems
remain unsolved.

The first one is to prove the existence of a phase transition in
the framework of equilibrium statistical mechanics. To this end,
one has to define an order parameter and a low-temperature phase
which is destroyed when the temperature increases. If one succeeds
in defining  an interface between the phases, the proof can be
based on contour arguments \cite{contour} which compare the
entropy and energy of the boundary of the low-temperature
clusters.

The second problem is the kind of the phase transition. According
to the Lee-Yang theory, it can be determined by evaluating
the asymptotic density of zeros around the real axis. However,
the fulfillment of this program needs consideration of much
larger systems than the ones considered here.

It is of interest also to construct the complete phase diagram of
the model in the $(z,w)$ plane, or in the $(x,y)$ plane, and to clarify
the relationship between the dynamics of the TASEP and the thermodynamics
of the effective two-dimensional model.

\section*{Acknowledgments}

The partial support by grant of the Representative Plenipotentiary
of Bulgaria to the Joint Institute for Nuclear Research in Dubna is gratefully
acknowledged. V. B. P. acknowledges the support by the RFBR grant No. 03-01-00780.

\section*{Appendix}

Here we present the derivation of expression (\ref{vstfin}) for the stationary
velocity in finite-size systems.

After summation over all final configurations $C$, Eq. (\ref{vs})
can be written
 as a sum over all clusters (connected sequences of neighboring particles)
in the initial configuration $C^0$:
\begin{equation}
v_{st}=\frac{1}{P}{L\choose{P}}^{-1} \frac{z}{1-z} \sum_{C^0}\sum_{i=1}^{m(C^0)}
\left(1-z^{N_i(C^0)}\right),
\end{equation}
where $m(C^0)$ is the number of clusters in configuration $C^0$
and $N_i(C^0)$ is the number of particles in the $i$-th cluster of
the configuration $C^0$. For further calculations, it is
convenient to introduce the following generating function of the
stationary velocity:
\begin{equation}
\label{Gam}
\Gamma (L,z,s)=\sum_{P=1}^{L-1}P {L\choose{P}} v_{st}(L,P,z)s^P .
\end{equation}
It is readily seen that the stationary TASEP can be mapped on the Fortuin-Kasteleyn
representation of the one-dimensional Potts model with external field \cite{Fortuin}.
To this end, we introduce the ring graph $G$ with $L$ vertices located
at the centers of the bonds of the TASEP ring $R$. Let $G'$ be a subgraph of $G$,
such that two neighboring vertices are connected by an edge if and only if the
common site of the corresponding bonds of $R$ is occupied. Thus, we obtain a one-to-one
correspondence of the TASEP configurations on $R$ and the subgraphs of $G$. Note that
the graph $G$ itself and the edgeless subgraph correspond to the fully occupied $(P=L)$ and
the empty $(P=0)$ rings, respectively. The velocity generating function (\ref{Gam}) can be
represented as a sum over all subgraphs:
\begin{equation}
\label{Gam2}
\Gamma (L,z,s)=\frac{z}{1-z}\sum_{G' \subseteq G}s^{e(G')}\sum_{i=1}^{k(G')}
\left(1-z^{n_i(G')-1}\right) - \frac{z}{1-z}\left(1-z^{L-1}\right)s^L ,
\end{equation}
where $e(G')$ and $k(G')$ are correspondingly the number of edges and the number
of connected components of $G'$, $n_i(G')$ is the number of vertices in the $i$-th
connected component of $G'$. The last therm is a correction to the fully occupied case.
The partition function of the $q$-state Potts model in external field can be expanded
over the spanning subgraphs as follows:
\begin{equation}
\mathcal{Z}_{Potts}(q,v,w)=\sum_{G' \subseteq G} v^{e(G')}
\prod_{i=1}^{k(G')}\left[ (w+1)^{n_i(G')} + q - 1\right].
\end{equation}
Here $v$ and $w$ are the coupling and external field parameters, respectively.
If we denote
\begin{equation}
f(v,w)=\frac{1}{(w+1)^L} \left.\frac{\partial\, \mathcal{Z}_{Potts}}
{\partial q}\right|_{q=1},
\end{equation}
then the generating function (\ref{Gam2}) can be written as
\begin{equation}
\Gamma (L,z,s)=
\frac{z}{1-z}f(s,0)-
\frac{1}{1-z}f(s,\frac{1-z}{z})-
\frac{z}{1-z}\left(1-z^{L-1}\right)s^L.
\end{equation}
By using the exact expression for the Potts model partition function,
\begin{equation}
\mathcal{Z}_{Potts}(q,v,w)=(q-2)v^L+(\lambda^+)^L+(\lambda^-)^L,
\end{equation}
with
\begin{equation}
\lambda^\pm = \frac{1}{2}\left[q+2v+w+vw \pm \sqrt{(q+2v+w+vw)^2-4v(q+v)(1+w)}\right],
\end{equation}
after some simplifications we find
\begin{equation}
\Gamma (L,z,s)=L z s\frac{(1+s)^{L-1}-(z s)^{L-1}}{1 + s - z s}.
\end{equation}
With the aid of the above exact expression we obtain the explicit expression (\ref{vstfin})
for the stationary velocity.


\begin{references}
\bibitem{Lig}T. M. Liggett, {\it Interacting Particle Systems} (Springer, 1985).

\bibitem{Spohn}H. Spohn, {\it Large Scale Dynamics of Interacting Particles} (Springer, 1991).

\bibitem{Dhar}D. Dhar, Phase Trans. {\bf 9}, 51 (1987).

\bibitem{Gwa}L. H. Gwa and H. Spohn, Phys. Rev. Lett. {\bf 68}, 725 (1992).

\bibitem{DerEvMuk}B. Derrida, M. R. Evans, and D. Mukamel, J. Phys. A {\bf 26}, 4911 (1993).

\bibitem{JanLeb}S. A. Janovsky and J. L. Lebowitz, Phys. Rev. A {\bf 45}, 618 (1992).

\bibitem{DerMal}B. Derrida and K. Mallick, J. Phys. A {\bf 30}, 1031 (1997).

\bibitem{Derrida}B. Derrida, Phys. Rep. {\bf 301}, 65 (1998).


\bibitem{DerLeb}B. D. Derrida and J.L. Lebowitz, Phys. Rev. Lett. {\bf 80}, 209 (1998).

\bibitem{Schutz}G. M. Sch\"utz, J. Stat. Phys. {\bf 88}, 427 (1997).


\bibitem{Pr}V. B. Priezzhev, Phys. Rev. Lett. {\bf 91}, 050601 (2003).

\bibitem{Evans1}M. R. Evans, J. Phys. A {\bf 30}, 5669 (1997).

\bibitem{RSSS} N. Rajewsky, L. Santen, A. Schadschneider and
M. Schreckenberg, J. Stat. Phys. {\bf 92}, 151 (1998).

\bibitem{YL52}
C. N. Yang and T. D. Lee, Phys. Rev.  {\bf 87}, 404 (1952).

\bibitem{BE02}
R. A. Blythe and M. R. Evans, Phys. Rev. Lett. {\bf 89}, 080601 (2002).

\bibitem{BE03}
R. A. Blythe and M. R. Evans, Braz. J. Phys. {\bf 33}, 464 (2003).

\bibitem{BE04}
R. Brak and J. W. Essam, J. Phys. A {\bf 37}, 4183 (2004).

\bibitem{BJJK}
R.A. Blythe, W. Janke, D.A. Johnston, and R. Kenna, J. Stat.
Mech.: Theory and Exper. P06001 (2004);
cond-mat/0405314 (2004).

\bibitem{BB}
J. Brankov and N. Bunzarova, Phys. Rev. E {\bf 71}, 036130 (2005).

\bibitem{BDL} I. Bena, M. Droz, and A. Lipowski, arXiv: cond-mat/0510278
(2005).

\bibitem{ZRP}
M. R. Evans and T. Hanney, J. Phys. A {\bf38}, R195 (2005).

\bibitem{Peierls} R. Peierls, Proc. Cambridge Philos. Soc. {\bf
32}, pt. 3, 447 (1936).

\bibitem{contour} M. Biskup, C. Borgs, J. T. Chayes, L. J. Kleinwaks, and
R. Kotecky, Phys. Rev. Lett. {\bf 84}, 4794 (2000).

\bibitem{Fortuin} P. W. Kasteleyn and C. M. Fortuin, J. Phys. Soc. Jpn. Suppl.
{\bf 26}, 11 (1969); C. M. Fortuin and P. W. Kasteleyn, Physica {\bf 57} 536 (1972).

\end{references}
\end{document}